\documentclass[aps,prb,amsfonts,amssymb,twocolumn,amsmath,preprintnumbers,nofootinbib,superscriptaddress,floatfix,showpacs]{revtex4}
\usepackage[dvips]{color}
\usepackage[dvips]{graphics}
\usepackage{graphicx}
\usepackage{bm}
\usepackage{amsmath}
\usepackage{amssymb}

\begin{document}



\title{Two regimes for effects of surface disorder on the zero-bias
conductance peak of tunnel junctions involving $d$-wave superconductors}

\author{M.~S.~Kalenkov}
\affiliation{P.N. Lebedev Physical Institute, Leninsky Prospect 53, Moscow
119991, Russia}
\author{M.~Fogelstr\"om}
\affiliation{Applied Quantum Physics, MC2, Chalmers,
S-41296 G\"oteborg, Sweden}
\author{Yu.~S.~Barash}
\affiliation{Institute of Solid State Physics, Chernogolovka, Moscow reg.
142432, Russia}
\affiliation{P.N. Lebedev Physical Institute, Leninsky Prospect 53, Moscow
119991, Russia}

\date{\today}


\begin{abstract}
Impurity-induced quasiparticle bound states on a pair-breaking surface of a
$d$-wave superconductor are theoretically described, taking into account
hybridization of impurity- and surface-induced Andreev states.
Further a theory for effects of surface disorder (of thin impurity surface
layer) on the low-bias conductance of tunnel junctions is developed.
We find a threshold $n_c$ for surface impurity concentration $n_S$,
which separates the two regimes for surface impurity effects on the
zero-bias conductance peak (ZBCP). Below the threshold, surface impurities
do not broaden the ZBCP, but effectively reduce its weight and generate
impurity bands. For low $n_S$ impurity bands can be, in principle, resolved
experimentally, being centered at energies of bound states induced by an
isolated impurity on the surface. For larger $n_S$ impurity bands are
distorted, move to lower energies and, beginning with the threshold
concentration $n_S=n_c$, become centered at zero energy. With
increasing $n_S$ above the threshold, the ZBCP is quickly destroyed in the
case of strong scatterers, while it is gradually suppressed and broaden in
the presence of weak impurity potentials. More realistic cases, taking into
account additional broadening, not related to the surface disorder, are also
considered.
\end{abstract}


\pacs{74.45.+c, 74.72.-h}


\maketitle



Inhomogeneities in anisotropically paired superfluids and
superconductors can noticeably modify properties of adjacent regions on
the scale of the coherence length. In particular, quasiparticle Andreev
bound states can arise in those regions near impurities, surfaces or
interfaces. Zero-energy Andreev bound states near impenetrable smooth
surfaces of $d$-wave superconductors are known to be the signature of a
sign change of the order parameter in a quasiparticle reflection event.
In high-temperature superconductors the zero-energy states are
identified, in particular, as resulting at low temperatures in the zero-bias
conductance peak (ZBCP) of NIS-junctions (see, for example, review articles
\cite{tan00,wendin01} and references therein).

In the present paper we report a theory for effects of surface
disorder on the ZBCP and, more generally, the subgap conductance. Our
starting point is the problem of quasiparticle bound states induced by an
isolated impurity, situated close or directly on $(110)$ surface
of a $d$-wave superconductor. Spectra of the bound states can be
found from the equation for the poles of the $t$-matrix, which takes
the form
\begin{equation}
{\rm det}\left(\hat{1}-V_{imp}\hat{\tau}_3
\hat{G}_0(\bm{r}_{imp},\bm{r}_{imp};\varepsilon)\right)=0 \enspace .
\end{equation}
Here $\bm{r}_{imp}$ is the impurity position and $V_{imp}$ the weight of the
potential $V(\bm{r})=V_{imp}\delta(\bm{r}-\bm{r}_{imp})$.
For simplicity, we disregard here the effect of the impurity on the
order parameter. The Nambu matrix retarded Green's function
$\hat{G}_0(\bm{r},\bm{r}';\varepsilon)$ describes the system in the
absence of the impurity. The quantity $\hat{G}_0(\bm{r}_{imp},
\bm{r}_{imp};\varepsilon)$ with identical first and second
coordinates is directly associated with the quasiclassical Green's
function averaged over the Fermi surface:
$\hat{G}_0(\bm{r}_{imp},\bm{r}_{imp};\varepsilon)=
N_f\left<\hat{\tau}_3\hat{g}_0(\bm{p}_f,\bm{r}_{imp};\varepsilon)
\right>_{S_f}$. Here we omitted the term, which should be eventually
included into the renormalized chemical potential; $N_f$ is the normal
state density of states on
the Fermi surface per one spin direction, $\hat{\tau}_3$ is the
respective Pauli matrix in particle-hole space. In the presence of
$(110)$ or $(100)$ surfaces in a $d$-wave superconductor, the averaged
off-diagonal components of the Green's function vanish
$\left<f_0(\bm{{p}}_f,x;\varepsilon)\right>_{S_f}=\left<f^+_0(
\bm{{p}}_f,x;\varepsilon)\right>_{S_f}=0$ throughout the
superconductor and for any $\varepsilon$. In
this case we obtain the following equation for the spectrum of
impurity bound states:
\begin{equation}
u\bigl<g_0(\bm{p}_f,\bm{r}_{imp};\varepsilon)
\bigr>_{S_f}=\pm\pi \enspace ,
\label{spectr}
\end{equation}
where dimensionless impurity potential is
introduced $u=\pi V_{imp}N_f$.  Near $(100)$ surface
the quasiclassical Green's function of a $d_{x^2-y^2}$-wave
superconductor does not depend on spatial coordinates
and takes the same form as in the bulk:  $g_0(\bm{p}_f,\bm{r};
\varepsilon)=-i\pi\varepsilon\left/\sqrt{\varepsilon^2-|
\Delta(\bm{{p}}_f)|^2}\right.$. Averaging this expression over a cylindrical
Fermi surface and substituting into Eq.~\eqref{spectr} leads to well
established results on energies of impurity bound states in $d$-wave
superconductors \cite{balatsky95,balatsky96}. Qualitatively, a well
pronounced low-energy resonance forms in the case of sufficiently large
strength of the impurity potential, for instance, close to the unitarity
limit. No resonance is present in the case of weakly scattering impurity,
when subgap impurity states lie mostly at higher energies, being strongly
broaden.

We demonstrate below, that impurity states near $(110)$ surface have
quite different structure as compared with the $(100)$ case, as an effect
of a hybridization of impurity- and surface-induced Andreev states.
In the presence of $(110)$ surface the quasiclassical Green's function
of a $d_{x^2-y^2}$-wave superconductor is spatially dependent and has a pole
at zero energy, associated with zero-energy surface Andreev bound states.
For this reason the spectrum of impurity bound states becomes dependent on
the distance of the impurity from the surface. Assume, for simplicity, that
the impurity is situated quite close or directly on $(110)$ surface, so
that the zero-energy surface states have strong influence on the formation
of the impurity states. In contrast with the $(100)$ case, a well pronounced
low-energy impurity resonance $\varepsilon_{imp}=\varepsilon'_{imp}+
i\varepsilon''_{imp}$,\,\, $|\varepsilon''_{imp}|\ll|\varepsilon'_{imp}|$
takes place in this case only for weakly scattering impurity potential. At low
energies the pole-like term dominates the Green's function and in the first
approximation $\left<g_0(\bm{p}_f,x=0;\varepsilon)\right>_{S_f}=\frac{r}{
\varepsilon+i0}$, where $r=\pi\bigl<\bigl|\tilde{\Delta}(\bm{p}_f)\bigr|
\bigr>_{S_f}$ is the residue of the pole-like term averaged over the Fermi
surface. The quantity $\bigl|\tilde{\Delta}(\bm{p}_f)\bigr|$ coincides with
the modulus of the bulk order parameter, if one disregards the surface
pair-breaking. In general, it is a normalized integral over the superconductor,
containing a spatially dependent (self-consistent) order parameter
\cite{bs97,barash00}. Inserting the pole-like
term into the equation for impurity bound states, we find the energy
of the resonance for weakly scattering impurity potential ($u\ll 1$):
$\varepsilon'_{imp}=\pm ur\left/\pi\right.$. In order to find the decay rate,
one should keep also the first low-energy term in the imaginary part
of the Green's function, averaged over the Fermi surface.
For nonzero low energies, narrow vicinities near the nodes of the order
parameter dominate the averaging of the imaginary part.
Assuming a cylindrical Fermi surface, we find $\left<g_0(\bm{p}_f,
x=0;\varepsilon)\right>_{S_f}\approx\frac{r}{\varepsilon+i0} -i\pi
\frac{|\varepsilon| }{\Delta_1}$, where $\Delta_1$ is the slope of the order
parameter at the node, taken in the bulk: $\Delta_1 = |\Delta'(
\varphi_{node})| $. With this low-energy expression for the averaged Green's
function we obtain from Eq.(\ref{spectr}) for $u\ll1$:
\begin{equation}
\varepsilon_{imp}= \varepsilon'_{imp} +
i\varepsilon''_{imp} = \pm ur/\pi- i r^2 u^3/(\pi^2\Delta_1) \enspace .
\label{impss}
\end{equation}
Self-consistent calculations give for $(110)$ surface orientation
$r\approx\Delta_0$ with a good accuracy, where $\Delta_0$ is the maximum value
of the bulk order parameter $\Delta(\bm{p}_f)=\Delta_0\sin{2\varphi}$.
Impurity-induced subgap states in $d$-wave superconductors have Andreev origin.
An imaginary part of the energy $\varepsilon''_{imp}$ is associated with an
escape of the quasiparticle from the impurity into the bulk of the superconductor.
This can take place due to impurity scattering, if for an acquired momentum
direction the quasiparticle energy exceeds the modulus of the $d$-wave order
parameter. For sufficiently small energy $|\varepsilon'_{imp}|\ll\Delta_0$ a
quasiparticle can run away from the impurity only with momenta in a narrow
vicinity of nodal directions of the order parameter. Then a well pronounced
impurity-induced quasiparticle resonance arises. For a larger subgap energy
of impurity-induced state, the escape can occur in a wider region of momentum
directions and the resonance is more broaden. As seen from Eq.\eqref{impss},
the resonance is well defined for weakly scattering impurity potential $u\ll1$.
In the opposite case of large strength of impurity potential $u\gtrsim 1$, the
impurity states formally approach $|\varepsilon'_{imp}|\sim\Delta_0$ and the
resonance itself is ill defined, because
$|\varepsilon'_{imp}|\sim|\varepsilon''_{imp}|$.

Similar to the impurity in the bulk of $d$-wave superconductor \cite{balatsky96},
the local density of states (LDOS) at the impurity site near $(110)$ surface
manifests the particle-hole asymmetry: for repulsive impurity potential a
resonance peak in the LDOS at this site takes place only for $\omega<0$,
while for attractive potential at $\omega>0$. This property is directly
associated with vanishing anomalous Gor'kov Green's function, describing
the $d$-wave superconductor with $(110)$ surface in the absence of the
impurity and taken at one and the same site
$F_0(\bm{r}_{imp},\bm{r}_{imp},\varepsilon)=
N_f\left<f_0(\bm{p}_f,\bm{r}_{imp};\varepsilon)\right>_{S_f}=0$.

The resonance energy is a function on the distance $x_{imp}$ between
the impurity and the $(110)$ surface. At sufficiently large distances
$x_{imp} \gg \xi_0, u\xi_0$ and not necessarily for weak scatterers,
we obtain $\varepsilon'_{imp}=\pm \dfrac{u{\rm
v}^2_f}{2\pi\Delta_1x^2_{imp}}$, $\varepsilon''_{imp}=-\dfrac{3u^3{\rm
v}^4_f}{ 8\pi^2\Delta^3_1 x^4_{imp}}$. The weight of the
corresponding peak in the LDOS at the impurity location diminishes
$\propto x^{-2}_{imp}$ with increasing $x_{imp}$, similar to
the weight of the zero-energy peak originated from the surface bound states.

It turns out that changes in the LDOS due to the impurity resonance, take
place entirely at the expense of the zero-energy surface states. The presence
of a single impurity near $(110)$ surface induces four
quasiparticle impurity states: two spin degenerated states with positive and
two with negative energies. Their contribution to the integral DOS
is 2, since each spin-polarized Andreev state with nonzero energy
contribute one half to the integral DOS \cite{chtchelk03, bardeen69}.
In its turn, the number of the zero-energy Andreev surface states in the
presence of one impurity reduces exactly by two (by one per spin
polarization), as compared with the case of no surface
defects. Thus, the total number of the surface and the
impurity quasiparticle states keeps constant. This is valid, actually,
for any number of impurities at (or near) the surface. The resonance
energy and decay rate depend also on surface-to-crystal orientation. The
above consideration can be easily generalized, with minor modifications,
to a wide range of the orientations. Only for surface orientations very close
to $(100)$, when the residue of the zero-energy pole-like term becomes
negligible (due to a small fraction of trajectories where the zero-energy
surface states take place), a small low-energy term
$-{\frac{2\varepsilon}{\Delta_0}}\ln\left(\frac{4\Delta_0}{-i\varepsilon}\right)$
dominates the averaged Green's function. As a result, the spectrum of
impurity bound states on $(100)$ surface coincide with the case of
impurity-induced states in the bulk of a $d_{x^2-y^2}$-wave superconductor.

Consider further the low-energy LDOS and the low-bias
tunnel conductance for the $d$-wave superconductor with $(110)$
surface in the presence of many impurities. Since the conductance is
sensitive mostly to impurities in the vicinity of the surface, we
introduce below a thin surface layer of
strongly disordered normal metal, which differs from the adjacent
only by its very short mean free path $l$. We consider a clean
$d$-wave superconductor in the half-space $x>d$. The surface layer
with impurity concentration $n_{imp}$ is placed at $0<x<d$ between a
smooth impenetrable specularly reflecting surface at $x=0$ and a
clean superconductor at $x>d$. There are no additional potential
barriers at $x=d$, between the disordered layer and the superconductor.
The thickness $d$ is assumed much less than all other characteristic
scales in the problem. The model is very similar to the thin dirty
layer model (TDL) \cite{cul84} except that it incorporates the
self-consistent $t$-matrix approach, allowing for
arbitrary strength of quasiparticle scattering by impurities, while the
TDL considers only the Born scattering. The other difference is that
real surface layer of normal metal with isotropic impurities is
introduced here, while special anisotropic scatterers
(Ovchinnikov impurities) are used in the models of surface roughness
\cite{ovchin69,cul84}. Our results, presented below, can be easily
reformulated for the case of Ovchinnikov impurities as well.

The quasiclassical $2\times2$
particle-hole matrix retarded Green's function $\hat{g}(\bm{p}_f,x;
\varepsilon)=\left(\begin{array}{cr} g & f \\ f^+ & -g \end{array}
\right)$ describes quasiparticle excitations and obeys Eilenberger's
equations. For thin disordered surface layer with the impurity self-energy
$\hat\sigma_{imp} (x;\varepsilon)\approx\hat\sigma_{imp} (0;\varepsilon)$,
the surface impurity concentration $n_S=n_{imp}d$ and the surface impurity
self-energy $\hat\zeta(\varepsilon)=2i\hat\sigma_{imp}(
x=0;\varepsilon)d$ enter the final results.
Off-diagonal elements of $\hat\zeta(\varepsilon)$ and
$\left<\hat{g}(\bm{{p}}_f,x;\varepsilon)\right>_{S_f}$ vanish in the case
of $(110)$ surface, so that  $\hat\zeta(\varepsilon)=\zeta(\varepsilon)
\hat\tau_3$.

The shape of the zero-energy peak in the density of states depends,
to a great extent, on the surface self-energy $\zeta(\varepsilon)$. The
surface self-energy, in its turn, is controlled by the Green's
function averaged over the Fermi surface.
We admit isotropic scatterers with arbitrary strength of potentials.
Introducing a dimensionless surface
impurity concentration $\tilde{n}_S=2n_{S}\bigl/(\pi\hbar{\rm v}_f N_f)$,
the surface self-energy parametrizes as:
\begin{equation}
\label{zeta}
\zeta(\varepsilon)=
\dfrac{i{\rm v}_f\tilde{n}_S}{\pi}\dfrac{\left<g(\bm{p}_f,0;\varepsilon)
\right>_{S_f}}{\dfrac{1}{u^2}-
\dfrac{1}{\pi^2}\left<g(\bm{p}_f,0;\varepsilon)\right>_{S_f}^2}\, .
\end{equation}
For the low-bias conductance only the Green's function on the surface
at low energies is of importance. The analytical solution we found for
the model gives for low energies
\begin{equation}
	\label{green:surf}
	g(\bm{p}_f,x=0;\varepsilon)=
	-i\pi \coth\left(
        \frac{\zeta(\varepsilon)}{\mathrm{v}_f |\cos\varphi|}
	-
	i\frac{\varepsilon}{|\tilde\Delta(\bm{p}_f)|}
        \right) \enspace .
\end{equation}
Eqs.~(\ref{zeta}), (\ref{green:surf}) were derived exploiting the
self-consistent spatial profile of the order parameter. Our further goal
is to solve them jointly in some important particular cases and describe
analytically the respective LDOS as a function of the energy, the
surface impurity concentration and the strength of impurity potential.
Also we represent below our fully
self-consistent numerical results for conductance spectra, based on
quasiclassical equations and boundary conditions for Riccati
amplitudes of the quasiclassical matrix Green's function
\cite{schopohl98,eschrig00}.

It turns out that the impurity layer is unable to broaden zero-energy
surface bound states, if surface impurity concentration is less than
the characteristic threshold: $\zeta(0)=0$ for
$\tilde{n}_S<\tilde{n}_c= \left<|\cos\varphi|\right>_{S_{f}}$. As we
found analytically, the delta-peak at zero energy occurs in the LDOS
for all surface impurity concentrations below the threshold $n_S<n_c$
(see, e.g., Eq. (\ref{bel})). Although, the weight of the peak is
quite sensitive to $n_S$ (see below). For the simplest model of a
quasi-two-dimensional superconductor with cylindrical Fermi surface
$\tilde n_c=2/\pi$. The value $\tilde{n}_c$ is determined by the
symmetry of pairing and does not depend on the particular spatial
profile of the order parameter and the form of its basis functions.
The threshold concentration of surface impurities exactly coincides
with the number $N_0(0)$ of the zero-energy bound states per unit
area of $(110)$ surface in the absence of the disordered surface
layer: $N_0(0)=n_c=(\pi/2)N_f{\rm v}_f \tilde{n}_c$. The effect is
associated with detaching the states from the zero-energy peak in
forming impurity bands.  To a certain extent, this bears an analogy
to what is known for impurity broadening of Landau levels and lifting
of their degeneracy in the two-dimensional electron gas in a strong
magnetic field \cite{entin78,dug03,skvortsov03}.  In the presence of
the disordered layer the number of zero-energy surface states
linearly depends on the surface impurity concentration $\tilde{n}_S$
and vanishes at the threshold:
$N_0(\tilde{n}_S)=N_0(0)(\tilde{n}_c-\tilde{n}_S)/\tilde{n}_c$.  The
threshold concentration remains equal to the number of the
zero-energy states on the clean surface also for arbitrary surface
misorientation angle $\alpha$, not only for $\alpha=45^\circ$. For
cylindrical Fermi surface ${n}_c(\alpha) = N_f{\rm v}_f\sqrt{
\displaystyle 1- \left|\cos2 \alpha\right|}$. It approaches its
maximal value for $\alpha=45^\circ$.

\begin{figure}[!tbh]
\centerline{\includegraphics[clip=true,width=3in,height=3in]{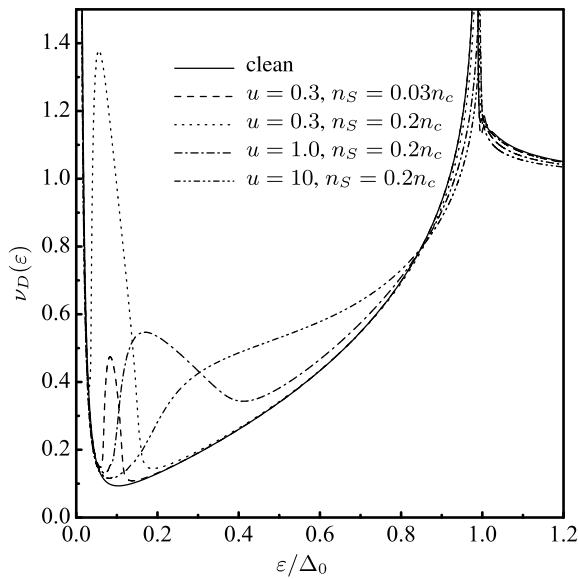}}
\vspace{0.3cm}
\caption{Tunnel density of states on $(110)$ surface as a function of
energy. Solid line represents tunnel density of states for a smooth clean
surface. Other curves describe effects of the surface impurity layer
with various strengths of impurity potentials and surface impurity
concentrations. Intrinsic broadening $0.002T_c$ is introduced
for resolving the zero-energy $\delta$-peak.}
\label{fig1}
\end{figure}

The behavior of the LDOS qualitatively differs below and above the impurity
concentration threshold. In the presence of the surface impurity layer
impurity bound states transform into two impurity bands, one with positive and
the other with negative energies. For low concentrations of the Born
impurities the bands are situated at low energies and the LDOS for the
impurity bands takes the form
\begin{equation}
	\label{imp:zone}
	\nu(\varepsilon)=
	\dfrac{2\nu_{max}}{W}
	\sqrt{(W/2)^2-(\varepsilon-\varepsilon_{imp}^{\prime})^2}
        \enspace .
\end{equation}
This expression is valid in the energy range, where the argument of the square
root function is positive. The characteristic bandwidth $W \sim \Delta_0 u
\sqrt{\tilde n_S}$ and the height of the LDOS in the center of the band
$\nu_{max}\sim \sqrt{\tilde n_S} / u$ are introduced in Eq.~\eqref{imp:zone}.
Numerical coefficients in expressions for $W$ and $\nu$ depend only on
$|\tilde\Delta(\bm{p}_f)|$. We omit here respective cumbersome
analytical formulas.
In a good agreement with self-consistent calculations, the center of the
Born impurity band is nothing but the position of the
impurity resonance on the $(110)$ surface:
$|\varepsilon'_{imp}|=u\bigl< | \tilde \Delta (\bm{p}_f) | \bigr>_{S_f}$.
As seen from these estimations, the impurity hump rises, becomes more narrow
and goes towards the zero energy, when the potential strength decreases.
Eq.~\eqref{imp:zone}, describing the shape of the impurity band in the
LDOS, is valid only for small surface impurity concentration
$u \nu_{max} \sim \sqrt{\tilde n_S} \ll 1$  and
implies that the width of the impurity band significantly exceeds
the width of separate impurity resonance $W\gg |\varepsilon''_{imp}|$.

Fig.~\ref{fig1} displays the energy dependence of the tunnel density of states
(i.e. the conductance at $T=0$) for clean $(110)$ surface, as well as for
surface disordered layers with various strengths of impurity potentials.
Here and below the momentum dependence of the transparency of NIS junction is
taken in the simple form $D( \varphi)=D_0 \cos^2 \varphi$, suitable for thin
and high potential barriers. Tunnel DOS and conductance are normalized to their
normal state values. At those energies, where the impurity peak stands out well
against the background, the tunnel density of states is simply proportional to
the LDOS on the surface with the energy-independent coefficient of the order of
unity. Eq.(\ref{imp:zone}) describes with a good accuracy impurity peaks
on two curves  with $u=0.3$ in Fig.~\ref{fig1}. The impurity peaks are situated
sufficiently close to the zero-energy peak. In the case $u=0.3$, $n_S=0.2n_c$
the impurity peak has a small asymmetric distortion, which is not taken into
account in Eq.(\ref{imp:zone}). With increasing $u$ the impurity bands become
less pronounced, more asymmetric and broaden. For sufficiently large $u$
impurity bands are strongly broaden at low $n_S$ over all subgap region.

\begin{figure}[!tbh]
\centerline{\includegraphics[clip=true,width=3in,height=3in]{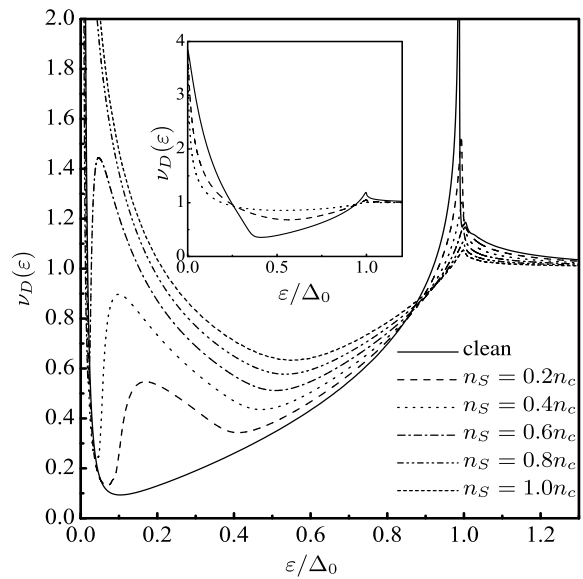}}
\vspace{0.3cm}
\caption{Tunnel density of states on $(110)$ surface as a function of energy,
taken for various surface impurity concentrations below and at the threshold.
The strength of the impurity potential is identical for all curves: $u=1$.
Inset shows the tunnel density of states for impurity concentrations above
the threshold: solid line -- $n_S = 2.5 n_c$, $u=0.3$, dashed line --
$n_S = 1.2 n_c$, $u=1$, dotted line -- $n_S = 1.07 n_c$, $u=10$.}
\label{fig2}
\end{figure}
Fig.~\ref{fig2} shows the evolution of impurity bands with varying the surface
impurity concentration below the threshold for the impurity potential $u=1$.
The impurity peak goes towards the lower energies and rises when the
concentration increases. Beginning with the threshold
concentration impurity bands are centered on zero energy, for any strength
of impurity potential. At the threshold concentration an impurity-induced
zero-energy peak in the tunnel DOS arises with a specific power-law spectral
shape $\nu_D(\varepsilon)\propto|\varepsilon |^{-1/3} $ in the low-energy
region $\ln(\Delta_0/|\varepsilon|) \gg 1$. This peak takes place for
arbitrary strength of impurity potential.

The weights of the ZBCP and the zero-energy peak in the LDOS, associated with
the surface zero-energy Andreev states, reduce with increasing $n_S$. Below the
threshold they vary linear in $(\tilde{n}_c-\tilde{n}_S)$, within the
logarithmic accuracy, and vanish at $n_S=n_c$. Close to the threshold
$0<\tilde n_c- \tilde n_S \ll \tilde n_c$
the low-energy momentum resolved LDOS
\begin{equation}
	\label{bel}
	\nu(\varphi,x=0;\varepsilon)=
	\dfrac{\pi| \cos \varphi |\,
	(\tilde{n}_{c}-\tilde n_S)}{
        \mathcal{S}_0
	\ln\left[
	\dfrac{
        \widetilde{\mathcal{S}} \mathcal{S}_0 \tilde\Delta_{max}
	}{
	\tilde n_c - \tilde n_S}
	\right]}\,
	\delta(\varepsilon).
\end{equation}
Here $\mathcal{S}(x) = \bigl\langle \cos^2 \varphi\ \delta (x - |\tilde
\Delta(\bm{p}_f) | ) \bigr\rangle_{S_f}$, $\mathcal{S}_0=\mathcal{S}(x\to+0)$,
$\mathcal{S}_0\ln\mathcal{\widetilde S} = \int_0^\infty  \mathcal{S'} (x) \ln(
\tilde\Delta_{max}/x ) dx$ and $\tilde\Delta_{max}$ is a maximal value of
$|\tilde \Delta (\bm{p}_f)|$ over Fermi surface.  For cylindrical
Fermi surface $\mathcal{S}_0= 2/\pi\Delta_1$. With spatially constant
order parameter we also find $\tilde{\cal S}=2$.
Eq.~\eqref{bel} is valid for low energies $|\varepsilon|\ll
(\tilde{n}_{c}-\tilde{n}_S)\Delta_{1},\ (\tilde{n}_{c}-\tilde{n}_S)^{3/2}
u\Delta_{1}$, where one can disregard the contribution from impurity bands.

Above the threshold $\tilde{n}_S>\tilde{n}_c$ the solution of Eqs.\eqref{zeta},
\eqref{green:surf} results in the following angular resolved surface
density of states at zero energy
\begin{equation}
	\label{nu1}
	\nu(\varphi,0;0)=
	\dfrac{
	\left(
	\dfrac{3\pi^2}{2u^2}
	+
	\ln\left[
	\dfrac{13.1}{\tilde n_S -\tilde n_c}
	\right]
	\right)^{1/2}
	}{\sqrt{6(\tilde n_S -\tilde n_c)}}
        |\cos\varphi | \enspace .
\end{equation}
This finite zero-energy value of the density of states, together with the
low-energy corrections (see, e.g., Eq.(\ref{dos:un})), describes the broaden
zero-energy peak above the threshold.
The condition $\zeta(0)/{\rm v}_f\ll 1$ implied in the derivation of
Eqs.(\ref{nu1}) applies not only close to the threshold. For sufficiently weak
scatterers, when the large term $3\pi^2/2u^2$ dominates the logarithmic
function, the expression for $\zeta(0)/{\rm v}_f$ remains also small for
$\tilde{n}_S \gg \tilde{n}_c$.  The Born approximation works in this case and
we find $\nu(\varphi ,x=0; \varepsilon=0)=|\cos \varphi|/(\rho
\tilde{n}_c)^{1/2}$, where $\rho=u^2\tilde{n}_S=d/l$. Here $l={\rm v}_f\tau$
and $\tau$ are the mean free path and the relaxation time in the disordered
layer. The inset of Fig.~\ref{fig2} demonstrates the behavior of the
low-energy tunnel density of states above the threshold. The heights of ZBCP,
which can be obtained from Eq.~\eqref{nu1}, are in agreement with those in
the inset of Fig.~\ref{fig2} within few percents.

Usually, the Born approximation is justified, if the first term in the
denominator of Eq.(\ref{zeta}) dominates the second one, i.e. the relation
$u^2|\left<g(\bm{p}_f,0;\varepsilon)\right>_{S_f}^2|/\pi^2\ll1$ satisfies.
The unitary approximation works in the opposite limit. In ordinary normal
metals $|\left<g(\bm{p}_f,0;\varepsilon)\right>_{S_f}^2|/\pi^2=1$ and the
condition for the Born (unitary) approximation to be valid reduces to the
standard form: $u^2\ll1$ ($u^2\gg1$). In superconductors
the quantity $|\left<g(\bm{p}_f,0;\varepsilon)\right>_{S_f}^2|/\pi^2$
can strongly deviate from unity at low energies.
In the absence of zero-energy (low-energy) states we have at low energies
$|\left<g(\bm{p}_f,0;\varepsilon)\right>_{S_f}^2|/\pi^2\ll 1$. This expands
the region of applicability of the Born approximation in superconductors
at low energies. However, in the presence of the zero-energy surface bound
states the pole-like term dominates the Green's function and takes very
large values on the surface at low energies. Moreover, since the
zero-energy states are dispersionless states, the Green's function, taken on
the boundary at low energies, remains to be quite large even after averaging
over the Fermi surface. Then, at sufficiently low energies we obtain
$|\left<g(\bm{p}_f,0;\varepsilon)\right>_{S_f}^2|/\pi^2\gg 1$. This condition
strongly modifies the behavior of the results, obtained in the Born and in
the unitary approximations \cite{pbbi99}. It is also seen, that
even though there are Born impurities in the disordered surface layer on the
normal metal sample ($u^2\ll 1$), in the case of the superconducting substrate
the presence of zero-energy surface states
places additional restrictions on admitting the Born approximation to
the same scatterers. They can scatter low-energy quasiparticles near the
surface with a large effective strength. This takes place, if the
impurity concentration is not sufficiently high and the zero-energy peak
in the LDOS is well pronounced. \cite{notes1} Under certain conditions the weak
impurity potentials $u^2\ll 1$ can be described even with unitary approximation
in the presence of the zero-energy surface states. This applies below the
threshold to describing the zero-energy peak and turns out to be very important
for our results. This is very different from various other circumstances where
the Green's function doesn't take large values and the Born approximation can
always be justified for sufficiently weak scatterers.

We find, that the Born approximation applies for describing sufficiently high
zero-energy peaks in the density of states, only if the surface impurity
concentration well exceeds the threshold surface concentration:
${n}_S\gg {n}_c$. \cite{note2} This condition can be quite restrictive.
A simple estimation gives $n_c\sim\hbar{\rm v}_fN_f\sim an_e$, $a$ being the
interatomic distance and $n_e$ the electron concentration.
For optimally doped $\mathrm{Y Ba_2 Cu_3 O_{7-\delta}}$ cut with $(110)$
surface, we evaluate the threshold surface concentration as
$n_{c} \approx1.3 \times10^{14}\mathrm{cm}^{-2}$. This is sufficiently large
(although well accessible) value of the threshold for the surface
concentration of point defects.

Conditions for the applicability of the Born and
the unitary approximations modify further for surface impurity concentrations
close to the threshold. The self-consistent analysis
of  Eqs.~\eqref{zeta}, \eqref{green:surf} shows that at $n_S\approx n_c$
the two large terms, which usually dominate the low energy behavior of
$\left<g(\bm{p}_f,0;\varepsilon)\right>_{S_f}$, almost cancel each other.
Then conventional criteria for the Born ($u\ll 1$) and the unitary
($u\gg 1$) approximations recover with a logarithmic accuracy.
In particular, this kind of conditions for the Born or the unitary
approximations arises in Eq.(\ref{nu1}), depending on what term
in the numerator dominates.

A characteristic feature of the zero-energy peak in the LDOS above the surface
concentration threshold, is a cusp-like shape of the peak. Indeed, the first
low-energy correction to the LDOS turns out to have a characteristic
energy dependence $\left<\delta\nu(\varphi, x=0; \varepsilon) \right>_{S_f}
\equiv\left<\nu(\varphi,0,\varepsilon)-\nu(\varphi,0,0)\right>_{S_f}
\propto|\varepsilon|$. The cusp-like shape of the broaden zero-energy peak
is clearly seen on the inset of Fig.~\ref{fig2} (see also Fig.~(6c) in
Ref.~\onlinecite{bbs97}, obtained with the thin dirty layer model and
Ovchinnikov impurities). While the zero-energy peak
in the LDOS originates from the sign change of the order parameter in a
quasiparticle reflection from the surface, the cusp-like behavior of the peak
comes from the narrow vicinities of the order parameter nodes.
Since for $(110)$ surface only the momentum directions close to the
surface normal are of importance, the tunnel DOS at low
energies acquires the same cusp-like behavior
$\sim|\varepsilon|$, although with a different numerical coefficient.
In the unitary limit, when the zero-energy density of states is described by
Eq.(\ref{nu1}) with $u \gg 1$, we get
\begin{equation}
	\label{dos:un}
	\left< \delta\nu(\varphi,0;\varepsilon) \right>_{S_f}=-
        \dfrac{\pi^2 |\varepsilon| \mathcal{S}_0}{8}
        \dfrac{\nu^4(0)}{\nu^2(0)(\tilde n_S-\tilde n_c)-4/(3\pi^3)}.
\end{equation}
This term becomes comparable with $\left< \nu(\varphi, x=0;0) \right>_{S_f} =
\nu (0)$ for $\varepsilon\sim \left(\tilde{n}_S-\tilde{n}_{c}\right)^{3/2}$.
Our estimations for sufficiently high and narrow ZBCP show that the width of
the peak manifests analogous dependence
$\gamma\sim\left(\tilde{n}_S-\tilde{n}_{c}\right)^{3/2}$.

In the Born limit we obtain $\left< \delta\nu(\varphi, x=0;\varepsilon)
\right>_{S_f}\approx-\pi|\varepsilon|/(4\rho\Delta_1)$. This
quantity is of the order of the zero-energy surface density of states $\sim
1/(\rho \tilde{n}_c)^{1/2}$ for $\varepsilon\sim\gamma\propto \rho^{1/2}$. For
the Born surface disordered layer $\zeta(0)/{\rm v}_f= \sqrt{\rho\tilde{n}_c}$.
It follows from here $\gamma\propto\sqrt{\rho}\propto \tau^{-1/2}$.  We note,
that the relation $\gamma\propto \tau^{-1/2}$ is a common feature of
the broadening of the zero-energy surface bound states by Born scatterers,
doesn't matter whether they are spread out as bulk impurities \cite{pbbi99},
occupy a surface layer with the thickness of the order or larger than the
coherence length \cite{pfs00}, or are collected only in a very thin surface
layer as it is the case in the present paper. The width of a well pronounced
impurity band turns out to be $\propto \sqrt{n_{imp}}$ also below the
threshold, as it is for $W$ in Eq.(\ref{imp:zone}), as well as in some
other circumstances \cite{hirsch93,kopnin02}.

\begin{figure}[!tbh]
\centerline{\includegraphics[width=85mm]{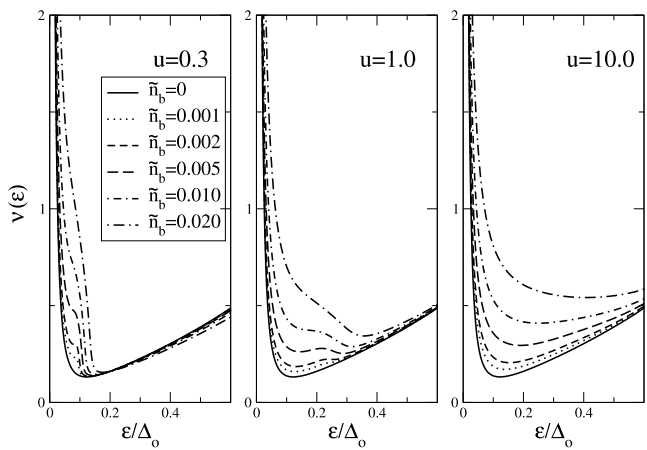}}
\vspace{0.3cm}
\caption{LDOS on $(110)$ surface as a function of energy, taken for
three values of strength of impurity potential $u$ and for various
bulk impurity concentrations. Dimensionless concentration of bulk
impurities $\tilde{n}_b$ is defined as $n_b=2\pi^2N_fT_c\tilde{n}_b$.
The lower curve corresponds to clean superconductor $\tilde{n}_b=0$.
With $\tilde{n}_b=0;\ 0.001;\ 0.002;\ 0.005;\ 0.01;\ 0.02$ one goes
from the lower to the upper curve on each panel.
Impurity humps are present for $u=0.3$, $u=1.0$ for a certain range of
concentrations $n_b$.}
\label{fig3}
\end{figure}

\begin{figure}[!tbh]
\centerline{\includegraphics[clip=true,width=3in]{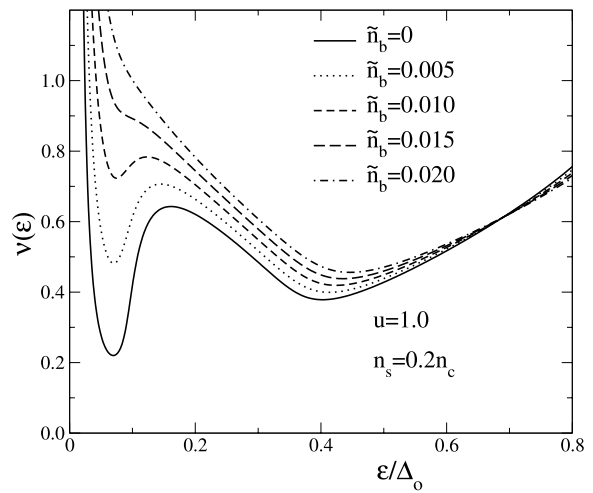}}
%
\caption{LDOS on $(110)$ surface in the presence of
bulk and surface impurities, taken for various bulk impurity
concentrations $\tilde{n_b}$, as a function of energy.
The strength of impurity potential $u=1$,
the surface impurity concentration $n_S=0.2 n_c$.}
\label{fig4}
\end{figure}

Consider further more realistic cases, taking into account also
some other reasons, not related to the surface disorder,
which can result in a broadening of the ZBCP in both regimes studied above.
Let, for example, impurities occupy not only the surface layer with
surface impurity concentration $n_S$, but also are spread out
over an extensive region of a superconductor as bulk impurities with small
bulk concentration $n_b$. Since energies of surface impurity states depend on
distance $x_{imp}$, in the presence of bulk impurities it is not obvious
whether impurity bands are fully smeared out and merge in a periphery of the
zero-energy peak, or they still survive, being centered
at energies of surface impurity states Eq.(\ref{impss}) for low $n_S$.
For making clear this question we have studied, firstly, the effect of bulk
impurities only. Fig.~\ref{fig3} shows the LDOS on $(110)$ surface for a
superconducting $d$-wave half-space with small bulk impurity
concentration $n_b$, which is constant everywhere up to $(110)$ surface.
As seen, surface impurity bands survive in the LDOS
for weak scatterers ($u=0.3$ and $u=1$) in a range of bulk impurity
concentrations $n_b$. Although, the bands are less pronounced and take place
in a more narrow region of $n_b$, as compared
with Figs.~\ref{fig1},~\ref{fig2} for the surface impurity layer model.
With increasing potential strength or/and impurity concentration $n_b$,
impurity bands are gradually smeared out and merge in the zero-energy peak.

\begin{figure}[!tbh]
\centerline{\includegraphics[width=3in]{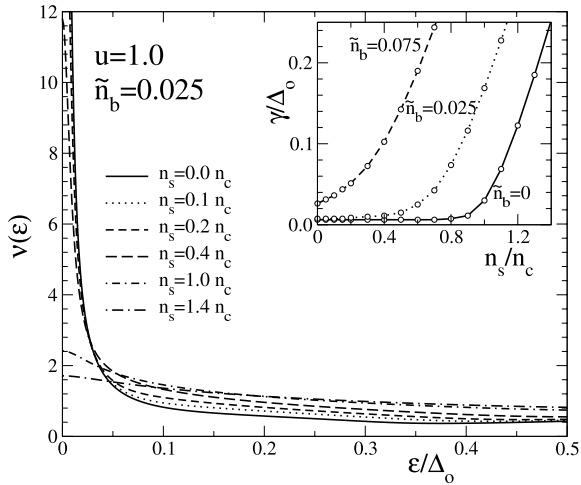}}
\vspace{0.3cm}
\caption{LDOS on $(110)$ surface in the presence of bulk and surface
impurities, taken for various surface impurity
concentrations $n_S$. The strength of impurity potential
$u=1$, the bulk impurity concentration $\tilde{n}_b=0.025$.
Inset shows the broadening $\gamma$ of the zero-energy peak as a function of
surface impurity concentration $n_S/n_c$, taken for various bulk impurity
concentrations $\tilde{n}_b$.}
\label{fig5}
\end{figure}

Effects of bulk impurities in the presence of thin surface
impurity layer on the LDOS on $(110)$ surface, are shown in Fig.~\ref{fig4}
for $n_S=0.2n_c$ and various $n_b$. Dimensionless concentration of bulk
impurities is defined as $n_b=2\pi^2N_fT_c\tilde{n}_b$. One can see,
that bulk impurities control the broadening of the zero-energy peak,
when surface disorder lies well below the threshold. The zero-energy peak and
the impurity bands fully merge into one broaden peak for comparatively small
concentration of bulk impurities $n_b\approx 0.04\pi^2N_f T_c$.
Even in this case the height of the zero-energy peak is
sensitive to the surface disorder, which controls relative weights
of central and peripheral regions of the broaden zero-energy peak.
The ``fine structure'' of the peak can be
resolved only for less $n_b$. Then well above the threshold concentration
the surface layer controls the broadening. Fig.~\ref{fig5} shows
the evolution of the low-energy LDOS with the surface impurity concentration
$n_S$ in the surface layer, in the presence of bulk impurities
($n_b=0.05\pi N_f T_c$).
The inset of Fig.~\ref{fig5} shows the broadening $\gamma$ of the zero-energy
peak as a function of surface impurity concentration $n_S/n_c$, taken for
various bulk impurity concentrations $\tilde{n}_b$.
The broadening $\gamma$ is the half of
the width of the zero-energy peak: $\nu(0)=2\nu(\gamma)$.
As seen in the inset of Fig.~\ref{fig5}, already at small concentrations
bulk impurities transform an abrupt change of the regimes of the broadening
at the threshold into a gradual crossover from one regime to another.

\begin{figure}[!tbh]
\centerline{\includegraphics[clip=true,width=3in,height=3in]{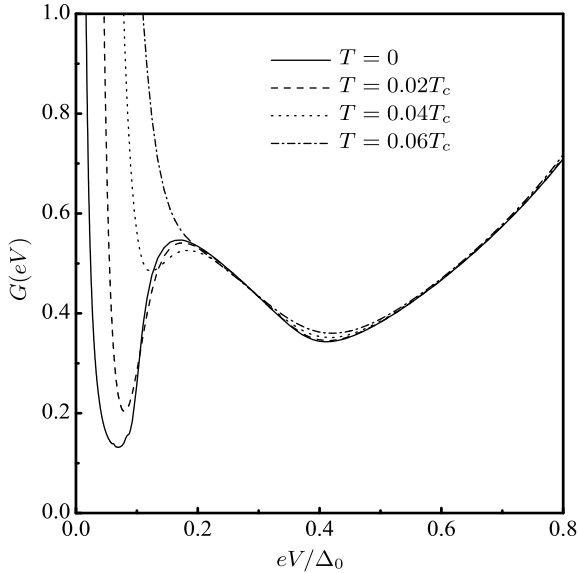}}
\vspace{0.3cm}
\caption{Tunnel conductance on $(110)$ surface for various temperatures as
a function of the applied voltage. With increasing the temperature the ZBCP
and impurity states broaden and strongly overlap. The strength of impurity
potential and the concentration are identical for all curves: $u=1$,
$n_S= 0.2n_c$.}
\label{fig6}
\end{figure}

\begin{figure}
\centerline{\includegraphics[clip=true,width=3in,height=3in]{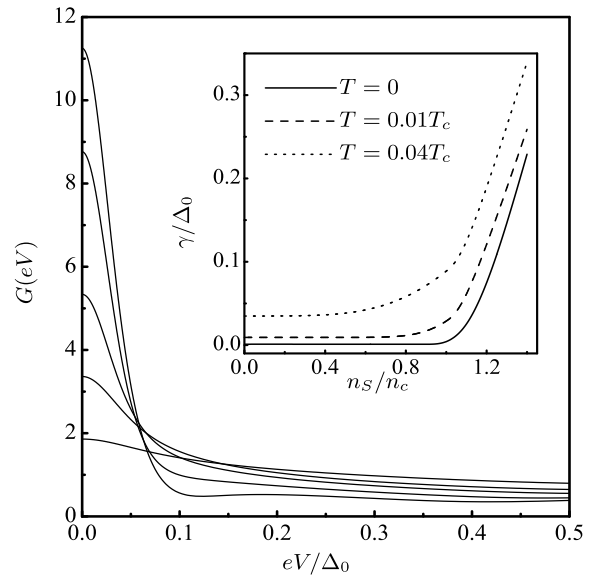}}
\vspace{0.3cm}
\caption{Tunnel conductance on $(110)$ oriented surface as a function of applied
voltage for different impurity concentration $n_S=0.2 n_c$, $n_S=0.4 n_c$,
$n_S=0.7 n_c$, $n_S=1.0 n_c$, $n_S=1.4 n_c$. Higher ZBCP corresponds to lower
concentration. Temperature and scattering impurity potential are the same
for all curves: $T=0.04T_c$, $u=1$. Inset shows the broadening $\gamma$ of
the ZBCP as a function of surface impurity concentration, taken for various
temperatures: solid line -- $T=0$, dashed line -- $T=0.01T_c$, dotted line --
$T=0.04T_c$. }
\label{fig7}
\end{figure}

Effects of finite temperatures also contribute to a broadening and
a suppression of the ZBCP, does not matter in the presence or absence
of surface disorder. The tunnel conductance as a function of the
applied voltage is shown on Fig.~\ref{fig6} for various temperatures.
The temperature broadening of both the zero-energy and the impurity
peaks takes place. The peaks fully merge into the one broaden peak
above a comparatively low temperature, which is $\approx0.06T_c$
for the particular case, considered on Fig.~\ref{fig6}.
The ``fine structure'' of the ZBCP arises below this temperature.
In the cases we consider, impurity bands result in two humps
(small satellites) whose heights are usually noticeably less than the ZBCP
in the self-consistent calculations. Only for concentrations very close to
the threshold the impurity peaks can become high and centered very close to
zero energy, while the zero-energy peak is already exhausted. We tried to
single out, whether this kind of ``spontaneous splitting'' of one broaden ZBCP
into two separate low-energy peaks can be observed below a characteristic
temperature in the case of high spectral resolution. Within the self-consistent
$t$-matrix approximation and for all sets of parameters we used, three peaks
always merge into one broaden peak before a ``spontaneous splitting''
described above could show up.

Fig.~\ref{fig7} shows the temperature broadening of the ZBCP for various
surface impurity concentrations. As seen in the range of low $n_S$, the
broadening
of the ZBCP, being fixed by the temperature, is almost insensitive to $n_S$.
At the same time the height of the peak noticeably decreases with
increasing $n_S$. The inset of Fig.~\ref{fig7} demonstrates that
effects of finite temperatures, analogously to the effect of bulk impurities,
transform the abrupt change of the
regimes at the threshold into a gradual crossover from one regime to
another. The surface impurity concentration can strongly influence
the width of the ZBCP only above the crossover region, while the
height of the ZBCP can be sensitive to the surface disorder for any
$n_S$.

Results of the present paper apply to tunnel junctions, when the
quasiparticle escape from the superconductor to the normal metal
does not have strong influence on the broadening of the zero-energy
states \cite{walker99,barash00}. For higher transparencies this mechanism
of the broadening becomes important. An information to what extent
impurities and effects of finite transparency are involved into forming
the broadening of the ZBCP in a particular NIS junction, in principle, could
be obtained experimentally, e. g. by measurements of the differential
shot noise at low voltages \cite{fog03}.

Thus, in the case of strong broadening of the ZBCP, induced by
sources, which are not related to the surface disorder,
impurity bands and the zero-energy peak, associated with Andreev
surface states, merge into one broaden ZBCP. This can be associated
also with the experimental resolution of the spectral weight.
Even in this case the height of the ZBCP is quite sensitive to the
surface disorder, whereas the width is not, since the disorder controls
relative weights of central and peripheral regions of the broaden ZBCP.
For a weak broadening and high experimental resolution
impurity bands can be identified, at least for low $n_S$ and $u$,
as low-energy humps near the zero-energy peak. When $n_s$
goes up, the bands become more close to the zero energy.
Then more resolution and less broadening are needed for
identifying them as separated from the zero-energy peak.
For concentrations $n_S\ge n_c$ impurity bands with positive
and negative energies merge into one impurity band,
centered at zero energy. Sources, not related to the surface disorder,
transform an abrupt change of the regimes of broadening
of the ZBCP, into a gradual crossover from one regime to another.
While the height of the ZBCP can be dominated by the surface disorder
for any $n_S$, its width is always independent of $n_S$ below the crossover
region. Above the crossover the surface disorder can strongly influence
the broadening of the ZBCP. It is worth noting, that special experimental
study of effects of surface disorder on the shape of the ZBCP has been
carried out in Ref.~\onlinecite{ap98}, where various degrees of the surface
quality have been prepared by changing ion irradiation
of YBa$_2$Cu$_3$O$_{7-\delta}$$\left/\right.$Pb junctions.
It was found in a certain range of irradiation doses, that the ZBCP width
is almost unchanged as a function of surface disorder, while the height
of the peak effectively decreases with increasing the irradiation effects.
This kind of behavior is in agreement with theoretical results of the
present paper.

We thank P.~Hirschfeld and N.~B.~Kopnin for useful discussions.
This work was supported, in part, by the Russian
Foundation for Basic Research under Grant No. 02-02-16643
(M.S.K. and Yu.S.B.) and by the Swedish Reseach Council (VR)(M.F.).


\begin{thebibliography}{999}
\bibitem{tan00}
S.~Kashiwaya and Y.~Tanaka,
Rep.~Prog.~Phys. {\bf 63}, 1641 (2000).
%
\bibitem{wendin01}
T.~L\"ofwander, V.~S.~Shumeiko and G.~Wendin,
Supercond.~Sci.~Technol. {\bf 14}, R53 (2001).
%
\bibitem{balatsky95}
A.~V.~Balatsky, M.~I.~Salkola, and A.~Rosengren,
Phys.~Rev.~B {\bf 51}, 15547 (1995).
%
\bibitem{balatsky96}
M.~I.~Salkola, A.~V.~Balatsky, and D.~J.~Scalapino,
Phys.~Rev.~Lett. {\bf 77}, 1841 (1996).
%
\bibitem{bs97} Yu.~S.~Barash and A.~A.~Svidzinsky,
JETP {\bf 84}, 619 (1997) [Zh. Eksp. Teor. Fiz. {\bf 111}, 1120 (1997)].
%
\bibitem{barash00}
Yu.~S.~Barash,
Phys. Rev. B {\bf 61}, 678 (2000).
%
\bibitem{chtchelk03}
N.~M.~Chtchelkacthev and Yu.~Nazarov,
Phys.~Rev.~Lett. {\bf 90}, art. 226806 (2003).
%
\bibitem{bardeen69}
J.~Bardeen, R.~K\"ummel, A.~E.~Jacobs, and L.~Tewordt,
Phys. Rev. {\bf 187}, 556 (1969).
%
\bibitem{cul84} F.~J.~Culetto, G.~Kieselmann and D.~Rainer in
{\it Proceedings of the 17th International Conference on
Low Temperature Physics}, edited by
U.~Eckern, A.~Schmid, W.~Weber and H.~W\"uhl
(North Holland, Amsterdam, 1984), \enspace p. 1027.
%
\bibitem{ovchin69} Yu.~N.~Ovchinnikov,
Sov. Phys. JETP {\bf 29}, 853 (1969) [Zh. Eksp. Teor. Fiz. {\bf 56}, 1590
(1969)].
%
\bibitem{schopohl98}
N.~Schopohl and K.~Maki,
Phys. Rev. B {\bf 52}, 490 (1995);
N.~Schopohl, cond-mat/9804064 (unpublished).
%
\bibitem{eschrig00}
M.~Eschrig,
Phys. Rev. B {\bf 61}, 9061 (2000).
%
\bibitem{entin78}
E.~M.~Baskin, L.~N.~Magarill, and M.~V.~Entin,
Zh.~Eksp.~Teor.~Fiz. {\bf 75}, 723 (1978)
[Sov.~Phys.~JETP {\bf 48}, 365 (1978)].
%
\bibitem{dug03}
A.~M.~Dyugaev, P.~D.~Grigoriev, and Yu.~N.~Ovchinnikov,
Pis'ma Zh.~Eksp.~Teor.~Fiz. {\bf 78}, 180 (2003)
[JETP Letters {\bf 78}, 148 (2003)].
%
\bibitem{skvortsov03}
I.~S.~Burmistrov and M.~A.~Skvortsov,
Pis'ma Zh.~Eksp.~Teor.~Fiz. {\bf 78}, 188 (2003)
[JETP Letters {\bf 78}, 156 (2003)].
%
%
\bibitem{pbbi99}
A.~Poenicke, Yu.~S.~Barash, C.~Bruder, V.~Istyukov,
Phys. Rev. B {\bf 59}, 7102 (1999).
%
\bibitem{notes1}
The presence of well pronounced zero-energy peak in the
surface LDOS does not inevitably mean the large height of the ZBCP.
For thick junctions only active tunneling trajectories, focused
within a narrow cone around the surface normal, contribute to the
conductance. This can strongly reduce the ZBCP (as well as the impurity
peaks), with respect to the background level.
%
\bibitem{note2}
Analogous condition can be obtained for scatterers, which are homogeneously
spread out as bulk impurities in a $d$-wave superconductor up to $(110)$
smooth surface, assuming that the zero-energy peak in the LDOS is sufficiently
high. In terms of results of Ref.~\onlinecite{pbbi99} we find in this case,
that the condition for the Born approximation to apply to low-energy
quasiparticle scattering by impurities near the surface takes the form
$n\gg N_f\Delta_0$.
%
\bibitem{bbs97}
Yu.~Barash, H.~Burkhardt, A.~Svidzinsky,
Phys. Rev. B {\bf 55}, 15282 (1997).\\
Since the TDL considers only the Born scatterers,
the condition $n_S\gg n_c$ concerns, in particular, the applicability
of the results for effects of the TDL on the low-energy spectra of $d$-wave
superconductors.
%
\bibitem{pfs00}
A.~Poenicke, M.~Fogelstr\"om, J.~A.~Sauls,
Physica B {\bf 284-288}, 589 (2000).
%
\bibitem{hirsch93}
P.~J.~Hirschfeld, N.~Goldenfeld,
Phys. Rev. B {\bf 48}, 4219 (1993).
%
\bibitem{kopnin02}
N.~B.~Kopnin,
Phys. Rev. B {\bf 65}, 132503 (2002).
%
\bibitem{walker99}
M.~B.~Walker and P.~Pairor, Phys. Rev. B {\bf 59}, 1421 (1999).
%
\bibitem{fog03}
T.~L\"ofwander, M.~Fogelstr\"om, and J.~A.~Sauls,
Phys. Rev. B {\bf 68}, 054504 (2003).
%
\bibitem{ap98}
M.~Aprili, M.~Covington, E.~Paraoanu, B.~Niedermeier, and L.~H.~Greene,
Phys.~Rev.~B {\bf 57}, R8139 (1998).
%
\end{thebibliography}
\end{document}